\documentclass{kluwer}
\usepackage[dvips]{graphicx} 
\begin{document}
\begin{article}
\begin{opening}
\title{A Survey Searching for the Epoch of Assembling of Hubble Types}            

\author{Du\'\i lia F. \surname{de Mello}}
\institute{Onsala Space Observatory, SE 43992 Sweden}                               
\author{Anna \surname{Pasquali}}
\institute{ESO/ST-ECF, Garching bei Muenchen, Germany}
\runningtitle{Hubble types}
\runningauthor{de Mello, Pasquali}

\begin{abstract} 
We have started a survey of galaxies at intermediate redshifts using 
the HST-STIS parallel fields. Our main goal is to analyse the morphology 
of faint galaxies in order to estimate the epoch of formation of the Hubble classification
sequence. The high resolution of STIS images (0.05$''$)
is ideal for this work and enable us to perform a morphological
classification and to analyse the internal structures of galaxies.
We find that 40\% of the 290 galaxies are early types and that there are 
more irregulars and ellipticals at the fainter magnitudes.
\end{abstract}
\keywords{galaxy evolution}
\end{opening}

\section{Introduction}

One of the key questions in galaxy evolution is the epoch of the
assembling of the Hubble types. How and when do galaxies acquire their
shape? Are ellipticals assembled in a monolithic collapse in the early
universe or do they form from subunits as predicted in the hierarchical
clustering scenario? In the latter scenario stars 
are formed continuously over a wide range of redshift and galaxies are 
assembled via many generations of mergers of smaller subunits 
(e.g. White \& Frenk 1991, Kauffmann et al. 1993). This prediction is strongly 
supported by the compilation of the cosmic star formation history (a.k.a. 
Madau plot) where no particular epoch of star formation is seen (Madau et al. 
1996, Blain 2001). This has frequently been used against the monolithic collapse 
theory which predicts rapid star formation at very high redshift (z$>$2) 
followed by a steep decline in the star formation rates 
(Eggen et al. 1962; Jimenez et al. 1998). However, in the Madau plot, 
the $\rho$$_{SFR}$ at a given redshift could be either due to massive galaxies 
experiencing modest bursts of star formation or newly formed dwarf galaxies. 
The distinction between these two types of galaxies is the main difference 
between the two models. One way to decide between the models is to search for 
their predictions. It is our goal to construct a database of galaxies at 
intermediate-{\it z} which will illustrate the galaxy population at that particular epoch. 
Therefore our database will be useful when searching for the predicted types of galaxies at 
intermediate-{\it z}.

\section{The Data and Results}

\begin{table} %
\begin{tabular}{lr|lr}                                        
\hline
STIS Field of View & 51$''$ $\times$ 51$''$ &Irregulars & 72 (24.8\%)\\
pixel size & 21$\times$ 21$\mu$m  & Spirals & 73 (25.2\%)\\
pixel plate scale & 0.05071$''$	  & S/Irr & 15 (5.2 \%)\\
Clear Filter& 2000\AA\ to 1$\mu$m	&    Irr/S & 9 (3.1\%)\\
Total Area covered & 69.4 arcmin$^{2}$  &  S0s & 45 (15.5\%)\\
Limiting Magnitude& 23.5		&    Ellipticals & 71 (24.5\%)\\
& & Mergers & 5 (1.7\%)\\
& & Total & 290\\
\hline
\end{tabular}
\caption[]{Survey Properties and Results}
\end{table}

We are using archive HST/STIS images obtained in parallel mode 
(PI Baum, ID 8549). Table 1 summarizes the Space Telescope Imaging Spectrograph (STIS) and
the Clear filter properties. The great advantages of these data are the high resolution and the
fact that the fields were randomly selected and were not particularly biased towards any environment. 
The exposure times were selected by the project that was executed when STIS 
was pointing in parallel. We used only the fields that have high signal-to-noise and
selected as extended sources all objects which were 2 $\times$ the stellar point spread function. We measured sizes down to 1 $\sigma$ of the sky 
level. Based on visual inspection for morphology classification a limiting magnitude of 23.5 was 
chosen. We used an area of the Hubble Deep Field (HDF) North observed with STIS to compare 
the Clear magnitudes with magnitudes obtained with WFPC2. We found that Clear magnitudes are 
1.053 magnitudes fainter than HDF magnitudes (de Mello \& Pasquali 2001) of the same galaxies. 
From the slitless spectra we have determined the redshifts of several objects in our sample.
Typical cases at z=$0.45$ are described in Pasquali \& de Mello (2002). We have identified 
290 galaxies brighter than the limiting magnitude and visually classified each galaxy 
according to their morphology. We used two qualifiers: early-type/late-type and
disk/no disk/irregular. First the galaxy was assigned either early-type or late-type. Then it 
was searched for the presence of a disk and assigned disk/no disk/irregular. 
The transition classes S/Irr and Irr/S reflect the strength of spiral feature. Galaxies which 
were peculiars (irregular) but had some sign of spiral arms were classified as Irr/S.
A seventh class was assigned to galaxies which had strong signs of past interaction, mergers.
Fig.1 shows a gallery of morphological types selected to illustrate typical galaxies of our
survey and Table 1 summarizes the results. We found that 40\% of our sample is made of early type
galaxies (E and S0) which is higher than what is found in the local universe (22\%) and in the Hubble 
Deep Field at intermediate-{\it z} (11\%, van den Bergh et al. 2000). This could be a sign of
evolutionary effect and/or to the fact that the HDF sample is too small (only 50 objects at z=0.25-0.6).
In order to check whether there is any evolutionary effect in our sample we have divided the 
sample according to magnitude intervals (Fig.2). We found that there are: (i) more 
irregulars and ellipticals at fainter magnitudes, (ii) spirals and S0s in all
magnitudes, and (iii) mergers only at fainter magnitudes.

\begin{figure}
\caption[]{Gallery of STIS images showing typical galaxies in our survey. 
Resolution is 0.05$''$ per pixel (see file attached).}
\end{figure}

\begin{figure}
\centerline{\includegraphics[angle=0,height=8.0cm,width=4.0cm]{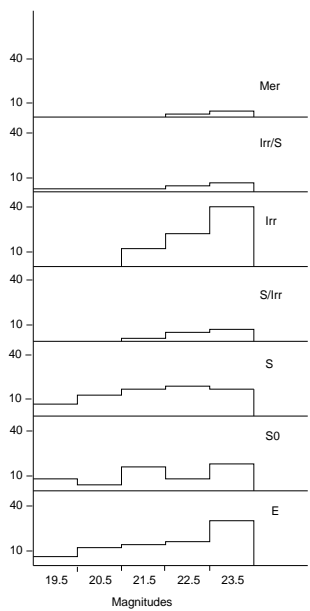}}
\caption[]{Number of galaxies of each morphological type versus magnitude}
\end{figure}

\end{article}
\end{document}